\documentclass{aa}  
\usepackage{graphicx}
\usepackage{txfonts}
\begin{document}
   \title{A Semi-Empirical Study of the Mass Distribution of Horizontal Branch Stars in M3 (NGC~5272)}
   
   \author{A. A. R. Valcarce 
   			   \and
           M. Catelan
           }
   
   \offprints{A. A. R. Valcarce}

   \institute{Pontificia Universidad Cat\'olica de Chile, Departamento de Astronom\'ia y Astrof\'isica,
              Av. Vicu\~na Mackena 4860, 782-0436 Macul, Santiago, Chile\\
              \email{avalcarc,mcatelan@astro.puc.cl}
             }

   \date{Received XXXXX XX, 2007; accepted XXXX XX, 2008}

  \abstract{}
   {Horizontal branch (HB) stars in globular clusters offer us a probe of the mass loss mechanisms taking 
   place in red giants. For M3 (NGC~5272), in particular, different shapes for the HB mass distribution have 
   been suggested in the literature, including Gaussian and sharply bimodal alternatives. 
   In the present paper, we study 
   the mass distribution of HB stars in M3 by comparing evolutionary tracks for a suitable 
   chemical composition and photometric observations.}
   {Our approach is thus of a semi-empirical nature, describing as it does the mass distribution that 
   is favored from the standpoint of canonical stellar evolutionary predictions for the distribution 
   of stars across the color-magnitude diagram. 
   More specifically, we locate, 
   for each individual HB star in M3, the evolutionary 
   track whose distance from the star's observed color and magnitude is a minimum. 
   We carry out 
   artificial tests that reveal that our method would be able to detect a bimodal mass distribution
   resembling the one previously suggested in the literature, if present.
   We also study the impact of different procedures for taking into account the evolutionary 
   speed, and conclude that they have but a small effect upon the inferred mass distribution. 
   }
   {
   We find that a Gaussian shape, though providing a reasonable first approximation, 
   fails to account for the detailed shape of M3's HB mass distribution. 
   Indeed, the latter may have skewness and kurtosis that deviate 
   slightly from a perfectly Gaussian solution. Alternatively,
   the excess of stars towards the wings of the 
   distribution may also be accounted for in terms of a bimodal distribution in which both the 
   low- and the high-mass modes are normal, the former being significantly wider than the latter. 
   }
   {However, we also show that the inferred distribution of evolutionary times is inconsistent
   with theoretical expectations. This result is confirmed on the basis of three independent sets 
   of HB models, suggesting that the latter 
   underestimate the effects of evolution away from the zero-age HB, and warning against 
   considering our inferred mass distribution as definitive. 
   }

               
\authorrunning{Valcarce \& Catelan}
\titlerunning{Mass Distribution of HB Stars in M3}

   \maketitle
%

\section{Introduction}

NGC~5272 (M3) is one of the best studied globular clusters in our galaxy. Extensive photometric 
studies of the cluster focusing on its variable stars were published as early as the beginning of 
the last century (Bailey 1913), and in spite of the many searches for stellar variability in the 
cluster field that have been carried out since (e.g., Roberts \& Sandage 1955; Szeidl 1973; Kaluzny 
et al. 1998; Corwin \& Carney 2001), until very recently previously unknown variable stars in the 
cluster field have continued to be reported in the literature (Clementini et al. 2004). 

Among M3's properties, of particular interest is its early classification, in terms of its RR Lyrae 
population, as an Oosterhoff (1939, 1944) type I (OoI) cluster. In fact, M3 is often used to define 
the properties of the OoI class (e.g., Smith 1995 and references therein). In spite of its role as 
the prototypical OoI cluster and its extensively studied properties, theoretical interpretation of 
the pulsation status of M3's RR Lyrae variables remains the subject of some controversy. In particular, 
Rood \& Crocker (1989) and Catelan (2004) have both called attention to the fact that canonical stellar 
evolution/pulsation theory appears somehow incapable of accounting for the sharply peaked distribution 
of (fundamentalized) periods that is found in the cluster (and possibly in other clusters as well). 
In particular, the Monte Carlo simulations presented by Catelan (2004), in which normal deviates 
were assumed [after Rood \& Crocker, who advocated a normal distribution for the masses of 
horizontal branch (HB) stars in M3], predicted period distributions were most often much 
flatter than observed. Among the possible solutions to the problem, these authors discuss the 
possibility of pulsation-induced mass loss leading to ``trapping'' of HB stars at some point in 
their evolution, as well as a rather peculiar {\em multimodal} mass distribution (see \S7 in 
Catelan 2004). 
 
Following up on the latter scenario, Castellani, Castellani \& Cassisi (2005) have recently argued, 
again on the basis of Monte Carlo simulations, that a sharply bimodal mass distribution may indeed 
be able to account for the observed period distribution. The shapes of their two ``period 
distribution-based'' mass modes turned out to be rather peculiar. More specifically, their mode 
responsible for the blue HB stars is essentially flat (between about $0.61\,M_{\odot}$ and 
$0.65\,M_{\odot}$), while the higher-mass mode responsible for the RR Lyrae and red HB stars 
corresponds to a Gaussian distribution with $\langle M \rangle = 0.68\,M_{\odot}$ and 
$\sigma_M = 0.005\,M_{\odot}$. Thus their suggested low- and high-mass modes are clearly 
detached from one another. Note, in addition, that Castellani et al. also advocate truncating 
the high-mass Gaussian mode at its mean mass value (i.e., retaining only the low-mass half of 
the Gaussian), in order to avoid the overproduction of red HB stars. 

As well known, the evolution of an HB star, for a given chemical composition, is primarily 
determined by its mass (or, more precisely, by the ratio between the envelope mass and the core 
mass; e.g., Caloi, Castellani, \& Tornamb\`e 1978). Therefore, in the case of relatively simple 
populations as monometallic globular clusters (such as M3 itself; e.g., Sneden et al. 2004), one 
expects, within the canonical framework,  
the detailed HB stellar distribution on the color-magnitude diagram (CMD) to reflect the 
underlying HB mass distribution. Accordingly, it should be possible, by careful examination of the 
available photometric data in such an extensively studied cluster as M3, to derive constraints on 
the shape of the underlying HB mass distribution. This is of relevance in the present context, 
since~-- as already discussed~-- the HB unimodality that was adopted as a working hypothesis in 
the HB simulations computed by Rood \& Crocker (1989) and Catelan (2004) has been identified as 
a possible explanation for the conflict between observed and predicted RR Lyrae period 
distributions in M3. In this sense, the main purpose of this paper is to check whether a 
multimodal mass distribution, as discussed by Catelan (2004), or a very peculiar bimodal mass 
distribution, as advanced by Castellani et al. (2005), is supported by the available CMD data.  

Unfortunately, not many studies have attempted to derive a mass distribution for M3 on the basis 
of the available CMD's. In this sense, while a normal distribution was favored by Rood \& Crocker 
(1989), and while other authors have succeeded in reproducing several different HB morphology 
parameters of the cluster without the need to invoke a multimodal mass distribution (e.g., Lee, 
Demarque, \& Zinn 1990; Catelan, Ferraro, \& Rood 2001b; Catelan 2004), data of much higher 
quality have become available in the literature since the late-80's for both variable and 
non-variable stars (e.g., Ferraro et al. 1997; Corwin \& Carney 2001), thus warranting a more 
detailed look into the photometric evidence for bimodality (or lack thereof) along the M3 HB. 
This is the main goal of the present paper. 

In \S2 we present the empirical data that were used in our study, along with the theoretical 
evolutionary tracks that were employed. In \S3 we describe the method we used, along with 
tests that demonstrate that this method would be able to detect a bimodal mass distribution 
resembling the one suggested in the literature, if one were indeed present. In \S4 we present 
the main results of our study. We close in \S5 by presenting a summary and conclusions.


\section{Data}

\subsection{Observational Data}

In order to reliably derive the mass distribution along the M3 HB, we need a photometric 
database containing both variable and non-variable stars. The main problem we face in this 
enterprise is related to the fact that, in order to derive reliable mean magnitudes and 
colors for the RR Lyrae stars, one needs time-series photometry covering the timespan of 
several days, and often weeks or months (particularly for variables with periods close to 
0.5~d), whereas non-variable red and blue HB stars can be much more straightforwardly 
measured. Fortunately, and as already stated (\S1), the M3 variables have been extensively 
studied in the literature, and reliable quantities for their ``equivalent static stars'' 
have recently been provided by Cacciari, Corwin, \& Carney (2005). 

For the variable stars, accordingly, we have used the Cacciari et al. (2005) database, 
which provides intensity-averaged magnitudes without the amplitude correction suggested 
by Bono, Caputo, \& Stellingwerf (1995), since the magnitude of the equivalent static star, 
according to the latter's models, are always within 0.02~mag of the intensity-mean value 
(see Marconi et al. 2003). For the average colors, in turn, we have used the Cacciari et 
al. static colors, which do include the Bono et al. amplitude-dependent corrections. 
Accordingly, the total number of RR Lyrae stars in our study is $133$, including $67$ 
fundamental-mode (RRab or RR0) pulsators with regular light curves and $43$ presenting 
the Blazhko effect (RR$_{\rm bko}$), plus $23$ pulsating in the first overtone (RRc or 
RR1 stars).

For nonvariable stars, in turn, we used the extensive photometric database by Ferraro 
et al. (1997), which includes $BVI$ data for around 45,000 stars. For stars in the outer 
regions of M3 ($r > 2\arcmin$), they used both CCD data obtained at the 3.6m CFHT telescope 
and data obtained using photographic plates (Buonanno et al. 1994), whereas for the inner 
regions {\em Hubble Space Telescope} (HST) photometry obtained with WFPC2 was provided. In 
this work we have adopted their $BV$ and $VI$ data for M3's outermost and innermost regions, 
respectively. As discussed in an Appendix, there may be a problem with the calibration of 
their $B$-band (photographic) data, so that we have effectively utilized only their $VI$ 
data to infer mass values for individual stars. 

\subsection{Evolutionary Tracks}

We used the set of (canonical) HB 
evolutionary tracks from Catelan et al. (1998a) for a chemical composition $Y_{\rm MS} = 0.23$
(envelope helium abundance on the zero-age main sequence), $Z = 0.001$. These are the same 
evolutionary models that were used in Catelan (2004), and they also form the basis for the 
recent calibration of the RR Lyrae period-luminosity relation by Catelan, Pritzl, \& Smith 
(2004). As a result, we have a total of 22 evolutionary tracks for masses between $0.496$ 
and $0.820 \ M_\odot$, extending from the zero-age HB (ZAHB) to core helium exhaustion. 

The possibility of a spread in helium abundances, as recently suggested by several different 
authors in the case of very massive/peculiar clusters, such as NGC~2808, NGC~5139 
($\omega$~Centauri), NGC~6388, and NGC~6441 (e.g., Caloi \& D'Antona 2007; Piotto et al. 2007), 
has not been taken into account in this study, due to the lack of empirical evidence supporting 
such a hypothesis in the specific case of M3; in particular, all the clusters for which a spread 
in $Y_{\rm MS}$ has been suggested present very long and well-populated blue HB ``tails,'' whereas 
such a feature is lacking in the case of M3.\footnote{After this paper had been submitted, 
a new preprint surfaced suggesting that a (relatively small) helium excess may be 
present among the blue HB stars in M3 (Caloi \& D'Antona 2008). Although not explicitly 
noted by those authors, the color distribution along the HB is well known to be degenerate 
in terms of second parameter candidates (e.g., Rood 1973), and so ``vertical'' information  
must be added in order to properly constrain the problem (e.g., Crocker, Rood, \& O'Connell 
1988; Catelan, Sweigart, \& Borissova 1998b; Catelan 2005). Accordingly, in a forthcoming study 
we will apply extensive such tests, using both spectroscopy and photometric diagnostics, to 
quantitatively constrain the extent to which helium may be enhanced among the cooler blue HB 
stars of M3 (Catelan et al., in preparation).} In the present paper, we shall accordingly 
restrict ourselves to the canonical scenario.

   \begin{figure}[t]
   \centering
      \includegraphics[width=9cm]{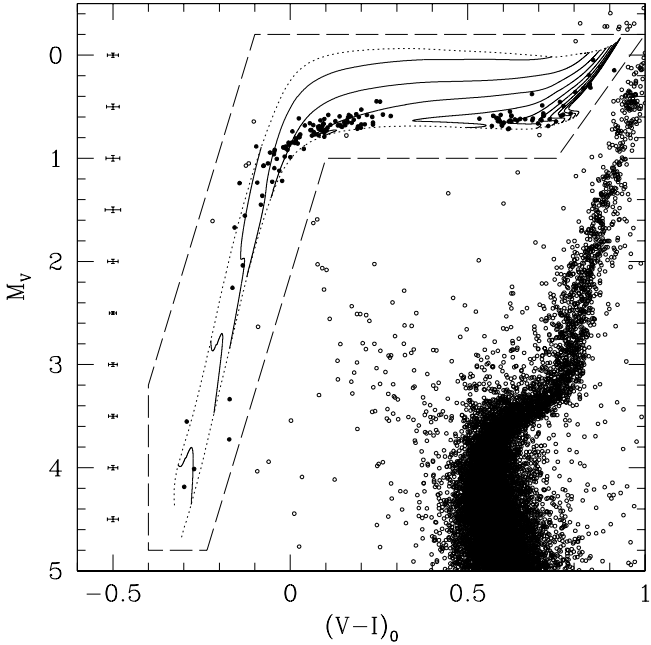}
      \caption{Evolutionary tracks for the adopted chemical composition ({\em solid lines}), 
	    along with the corresponding ZAHB and helium exhaustion loci ({\em dotted lines}), are 
		overplotted on the CMD for nonvariable stars ({\em open circles}) in the inner region 
		of M3. The evolutionary tracks correspond to masses from $0.5\, M_\odot$ ({\em left}) 
		to $0.8\, M_\odot$ ({\em right}), increasing in intervals of $0.025\, M_\odot$. The
		locus occupied by HB stars is schematically indicated by the dashed lines; stars within 
		this region plotted with {\em solid circles} have had mass values inferred using our 
		procedure.   As can be seen, the adopted distance modulus, $\mu_V=15.00$~mag, leads 
		to a nice agreement between the observed and predicted lower envelopes of the HB
		distribution in this diagram.}
         \label{CMDNoVar}
   \end{figure}

   \begin{figure}[t]
      \centering
      \includegraphics[width=9cm]{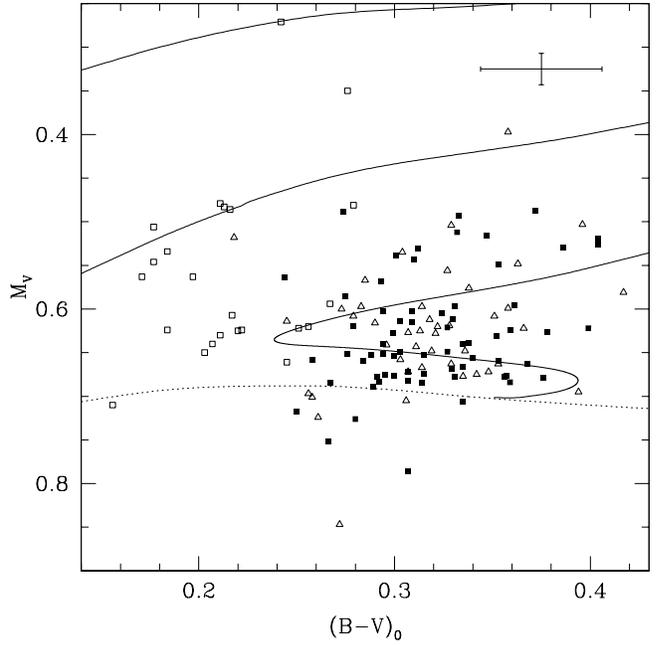}
      \caption{Evolutionary tracks for the adopted chemical composition ({\em solid lines}), 
	    along with the corresponding ZAHB ({\em dotted line}), are overplotted on the CMD for 
		M3 RR Lyrae variable stars. Different symbols are used for the RRab ({\em filled squares}), 
		RRc ({\em empty squares}), and RR$_{\rm bko}$ ({\em triangles}) stars. The solid lines, 
		from left to right, represent evolutionary tracks for $0.600$, $0.625$, and $0.650\, M_\odot$. 
		A distance modulus $\mu_V=15.00$~mag was again adopted.}
         \label{CMDVar}
   \end{figure}

\section{The Method}

The aforementioned evolutionary tracks are initially provided in the theoretical plane 
$[\log(L/L_\odot),\,\log(T_{\rm eff})]$, whereas the empirical data are in the form of 
broadband filters-based magnitudes and colors. Therefore, in order to be able to infer the 
masses of HB stars on the basis of their positions in a CMD, we have transformed the 
evolutionary tracks to the observational coordinates ($M_B, M_V, M_I$) on the basis of 
the color transformations and bolometric corrections by VandenBerg \& Clem (2003) for 
a metallicity ${\rm [Fe/H]}\simeq -1.5$ and abundance of the alpha elements 
$[\alpha/{\rm Fe}]\simeq +0.3$ (e.g., Sneden et al. 2004). Interpolation on their 
tables was carried out using the algorithm by Hill (1982). 

In order to increase the internal precision in the mass determination, 
we have generated additional 
evolutionary tracks, with a separation of $2\times 10^{-4} M_\odot$ between consecutive 
values, by interpolating (also using the Hill 1982 algorithm) on the original tracks 
available to us. Note that Hill's is a very powerful Hermite interpolation algorithm, 
which indeed proved of great assistance in dealing with the non-linearities that are 
observed in the shapes of the evolutionary tracks as a function of mass. 

Apparent magnitudes of M3 HB stars have been transformed to absolute magnitudes using a 
distance modulus $\mu_V = 15.00$~mag in the $V$ band. This is based on a comparison between 
our evolutionary tracks for $Z=0.001$ and the M3 observations: as well known, HB stars spend 
most of their lifetimes close to the ZAHB, and this is only consistent with our theoretical 
models for a distance modulus close to the indicated value. On the other hand, Harris (1996) 
provides $\mu_V=15.12$~mag instead; if this value were adopted, we would find the rather 
untenable result that more than 80\% of M3's HB stars would be in a very advanced evolutionary 
stage. We later discuss the effect of uncertainties in $\mu_V$ upon our results.

\begin{figure*}[t]
\begin{minipage}{0.5\linewidth}
\centering
\includegraphics[width=8.5cm]{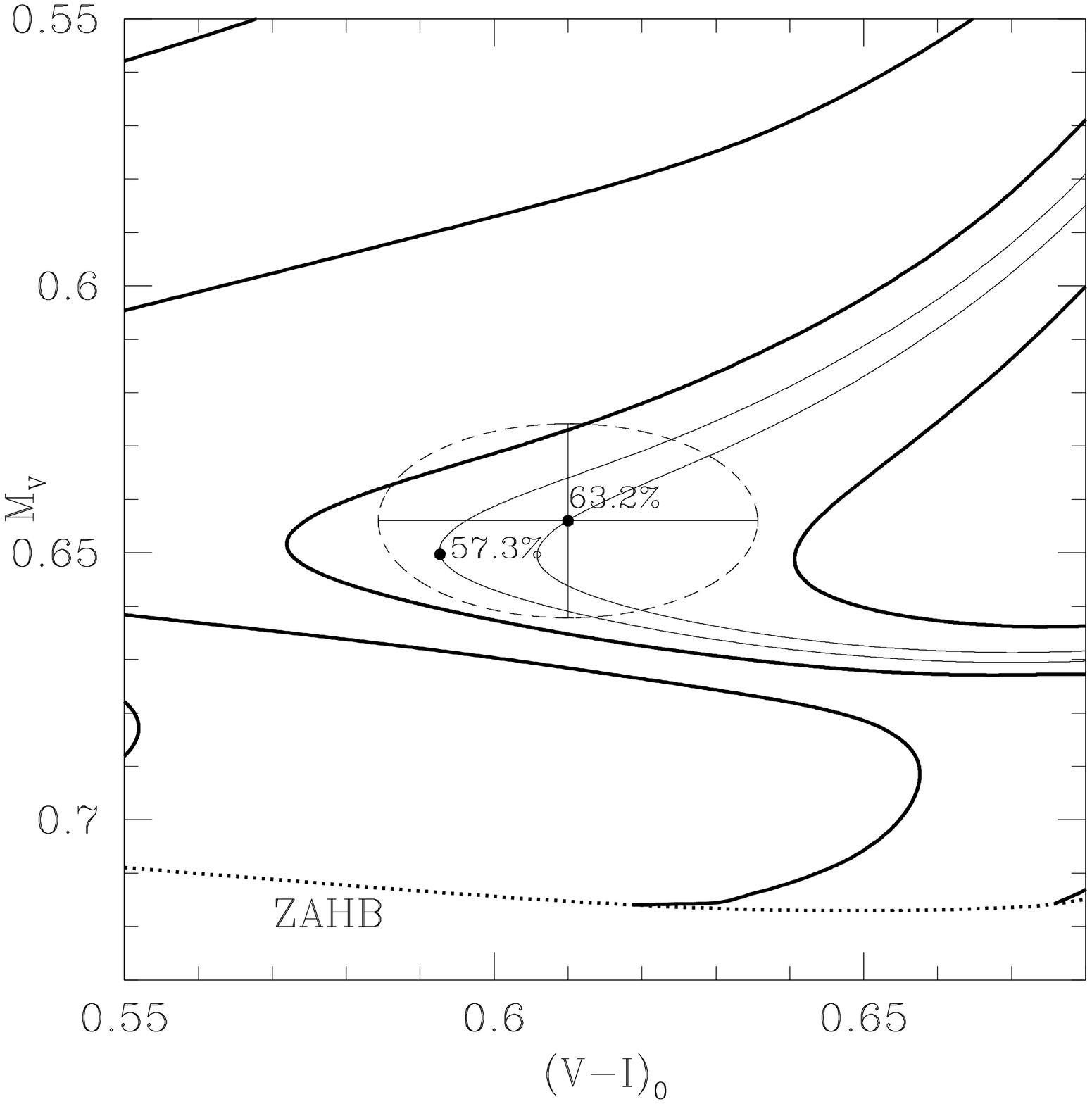}
\end{minipage}%
\begin{minipage}{0.5\linewidth}
\centering
\includegraphics[width=8.5cm]{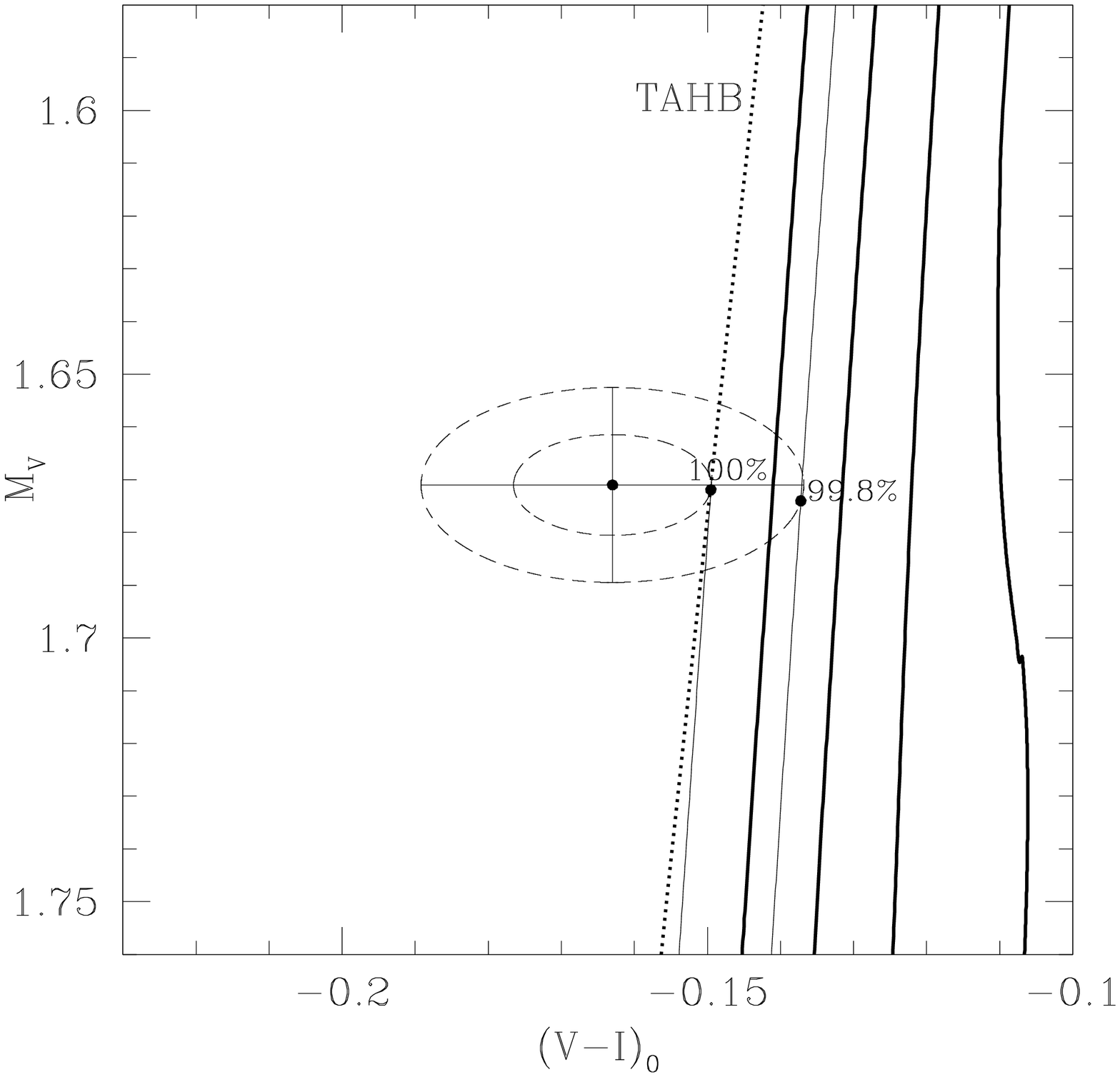}
\end{minipage}
\caption{Method used to determine the mass for each data point. {\em Left panel:} a red HB
	    star. The ZAHB is shown as a {\em dotted line}, whereas {\em bold lines} show evolutionary tracks for 
		$0.65, 0.66, 0.67$ and $0.68 \, M_\odot$ ({\em from left to right}). The ellipse is the isoprobability 
		contour for the 1-sigma error bars. The mass corresponding to the evolutionary track that passes 
		closest to the center of the ellipse ({\em thin solid line, right})
		is $0.6744 \, M_\odot$, whereas that corresponding to the minimum evolutionary speed
		({\em thin solid line, left}) is $0.6726 \, M_\odot$. The evolutionary lifetime $t_{\rm evol}$ 
		along the HB is also shown in both cases, given as a percentage of the total HB lifetime. 
		{\em Right panel:} a blue HB star. 
		{\em Bold lines} are evolutionary tracks for $0.545, 0.550, 0.555$ and $0.560 \, M_\odot$ 
		({\em from 
		left to right}). Here no HB track goes through the actual data point, and so we adopt for
        the star the mass corresponding to the track that comes closest to it, and still inside its 
		1-sigma error ellipse~-- in this case, 
		$0.541 \, M_\odot$, which corresponds to a star at He exhaustion ({\em thin solid line, left}). 
		The evolutionary track for the 
		mass value with the smallest evolutionary speed that still goes through the 1-sigma ellipse is 
		shown as a {\em thin solid line (right)}; this corresponds to a star with a mass of 
		$0.547 \, M_\odot$ that has completed 99.8\% of its HB evolution.} 
     \label{Elipses}
\end{figure*}

The colors were corrected for reddening using values for $E(B\!-\!V)$ and $E(V\!-\!I)$ of 
$0.010$ (Harris 1996) and $0.016$~mag (Rieke \& Lebofsky 1985), respectively. The adopted 
$E(B\!-\!V)$ value is only 0.003~mag smaller than the one implied by the Schlegel, 
Finkbeiner, \& Davis (1998) maps. 

We inferred the mass and the evolutionary times (given as number fractions, where 0\% 
corresponds to the ZAHB and 100\% to the helium exhaustion line [``terminal age HB,'' 
or TAHB]) for each individual HB star by choosing the track that most closely matched 
the star's observed CMD position (Figs.~\ref{CMDNoVar} and \ref{CMDVar}), with individual 
error bars being based on the observational (photometric) errors (see Fig.~\ref{Elipses}). 
We have also carried  
out tests in which the evolutionary times are explicitly taken into account in the mass 
derivation (\S\ref{sec:3MET}), 
but have not found important differences in the inferred mass distribution, at least 
in the case of M3-like HB morphologies, with respect to this more straightforward procedure.  

For the cases in which 
no individual error bars are provided in the original photometry, we have computed synthetic 
observational error bars using the method described by Catelan et al. (2001b; see their \S3.1), 
which invokes an exponential law for the errors in the observed magnitudes. The photometric 
error bars are of particular importance in the cases of stars lying close to the dashed 
line in Figure \ref{CMDNoVar}, since they define whether a star can reasonably be assigned 
to the HB phase or not (see also Fig.~\ref{Elipses}, {\em right panel}). 

 \begin{figure*}
	\centering
	\includegraphics[width=18cm]{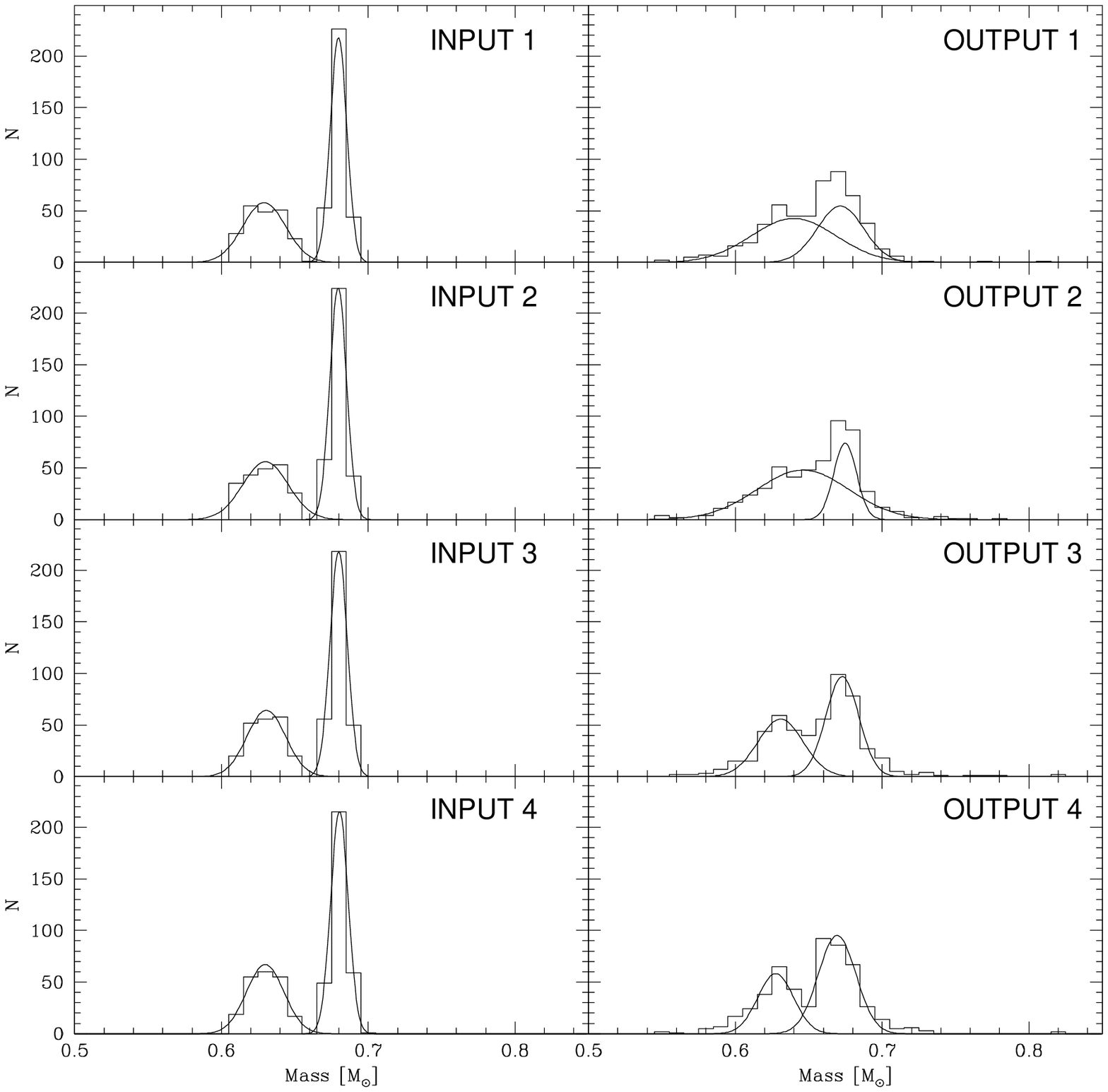}
	\caption{{\em Bimodality test}: 
        For four input bimodal mass distributions ({\em left panels}) we determined the 
		mass distribution using our method ({\em right panels}). Due to the observational 
		errors, we could not recover precisely the same input distributions, but in no case 
		has a unimodal mass distribution been derived on the basis of an input bimodal distribution.}
  	\label{BimTest}%
  \end{figure*}

	\begin{figure*}
  	\centering
    \includegraphics[width=18cm]{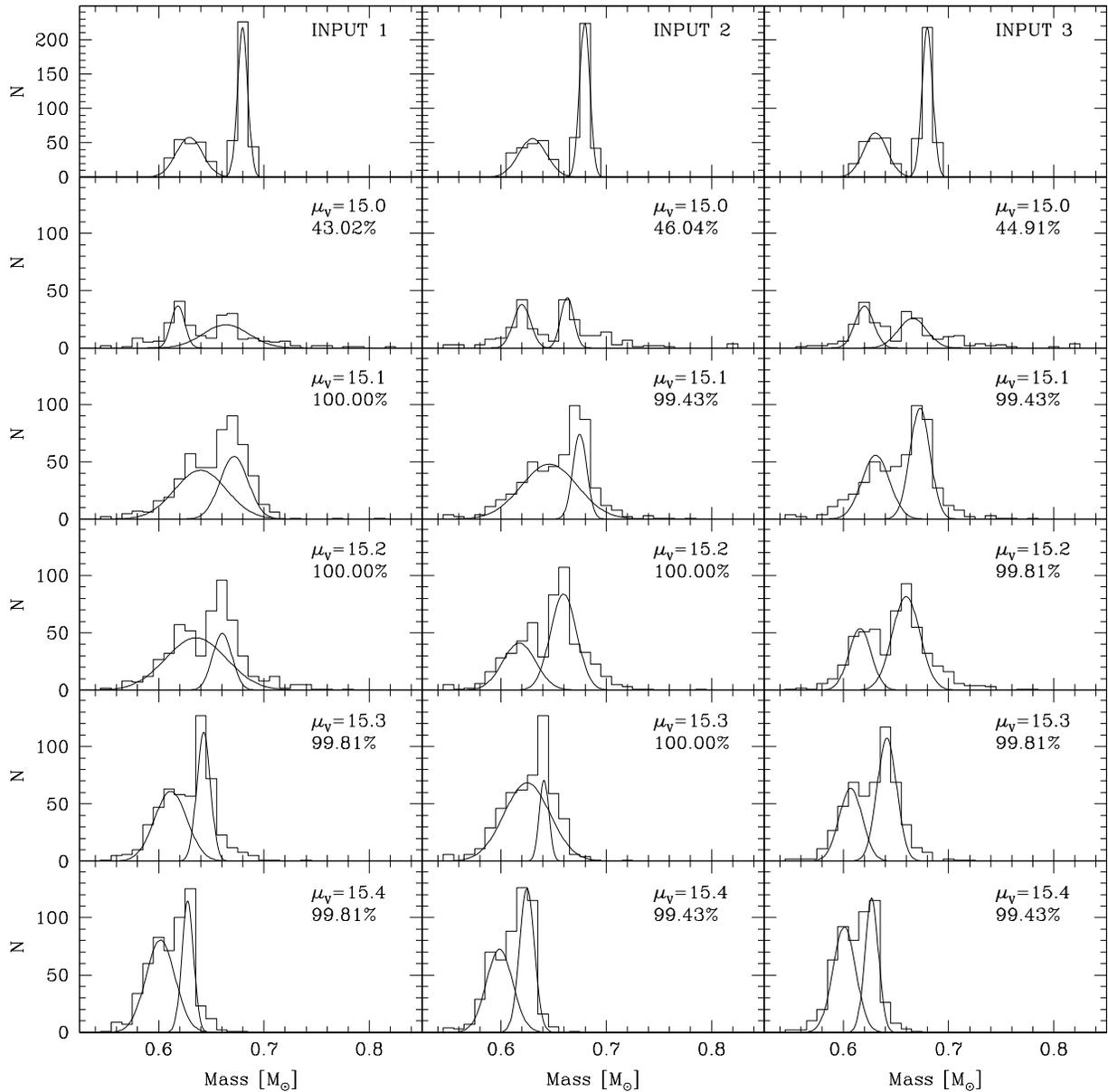}
    \caption{{\em Distance modulus test}: For three input bimodal mass distributions 
      ({\em upper panels}) and assuming different distance moduli (as indicated
      in the insets, along with the percentage of recovered HB stars), 
      we have derived the mass distribution using our method. Input distributions 1 to 3 are the 
      same as in the previous figure. As can clearly 
      be seen, we are always able to recover the input bimodality, albeit not
      without incurring some systematic errors in the placement of the mass peaks.  
    }
    \label{DMTest}
  \end{figure*}

  \begin{figure*}
  \centering
    \includegraphics[trim=10mm 65mm 10mm 5mm,clip=true, width=18cm]{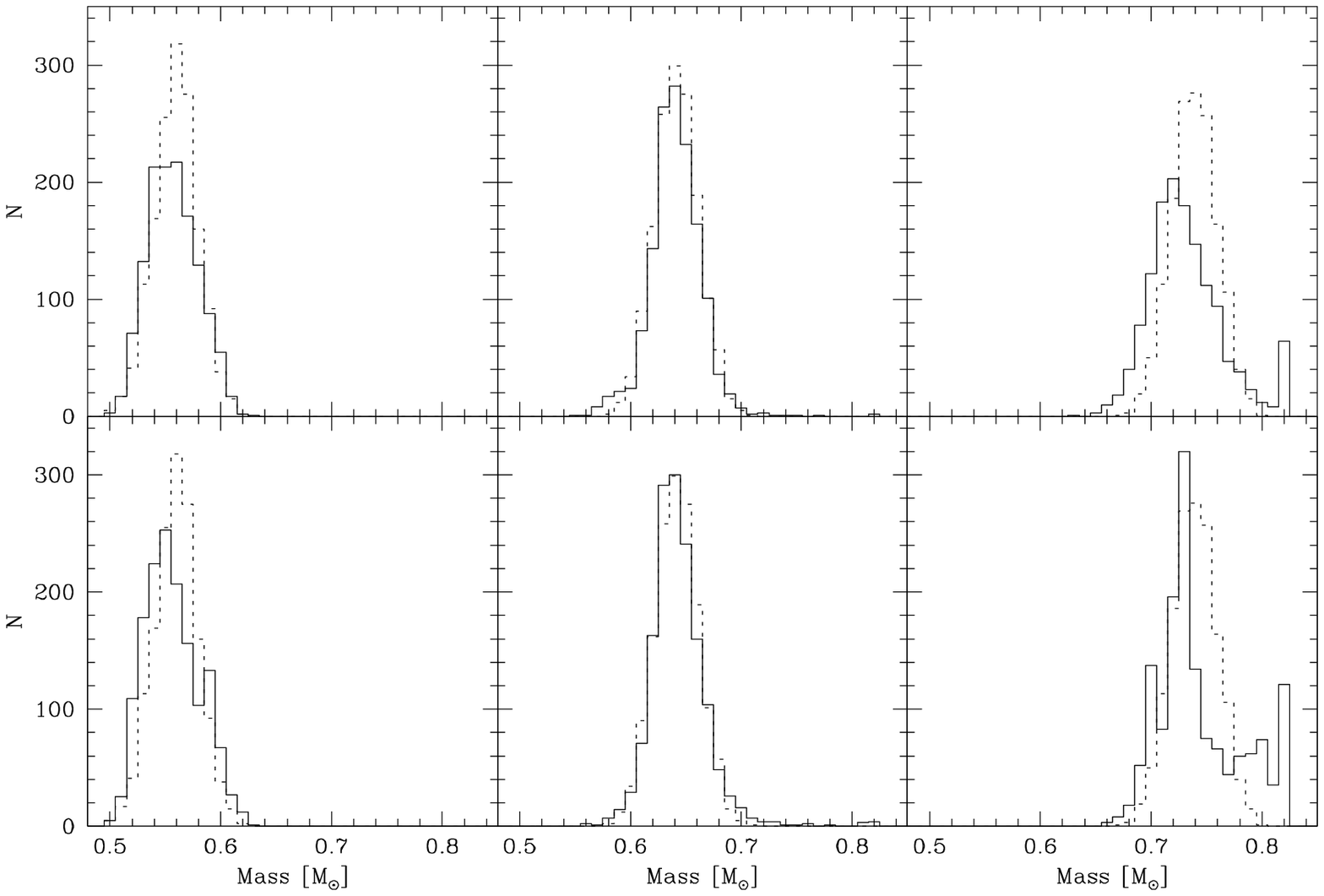}
    \caption{{\em Methodology test}: For known input mass distributions ({\em dashed lines})
      HB simulations with 1500 stars were computed as described in the text, and the mass 
	  distribution was then inferred ({\em solid lines}) from the resulting CMD distribution 
	  using two different methods: evolutionary tracks that pass closest to the individual 
	  data points ({\em upper panels}) 
	  and evolutionary track that passes through the 1-$\sigma$ 
	  error ellipse with the smallest evolutionary speed $v_{\rm evol}$ ({\em bottom panels}). 
	  On the {\em left and right panels} the cases of completely blue and completely red HB 
	  morphologies are shown, respectively, whereas the middle panel refers to the case of 
      a more even distribution along the CMD, as more appropriate in the case of M3. As can 
	  clearly be seen, while the present method becomes less reliable for extreme HB types, 
	  it does provide an excellent description of the input mass distribution for an M3-like 
	  HB morphology.  
    }
    \label{MDMethods}
  \end{figure*}

In this sense, for stars falling at positions above the TAHB or below the ZAHB (dotted lines in 
Fig.~1), we used these observational errors to determine, where applicable, the most likely 
mass value allowed for by the latter. We considered that magnitude and color errors 
($\sigma_{M_V}$ and $\sigma_{\rm color}$, respectively) are standard deviations of a normal 
distribution centered at the most probable value, which implies that a star has the same 
probability to have a mass value determined by the points located at 
$[M_V\pm\sigma_{M_V},\,(B\!-\!V)_0]$ or $[M_V,\,(B\!-\!V)_0\pm\sigma_{\rm color}]$. In fact, 
one has concentric ``isoprobability'' ellipses centered at $[M_V,\,(B\!-\!V)_0]$ (see 
Fig. \ref{Elipses}). The mass assigned to a star thus corresponds to the track in our grid 
of evolutionary tracks intersecting the error ellipse with the smallest semi-major axis. We 
used the ellipse with semi-major axis equal to the observational errors to assign errors to 
individual mass values ($\sigma_M$). Note, however, that some stars can fall much above the 
TAHB or below the ZAHB limits, implying that they are not bona-fide HB stars within their 
assigned photometric errors (i.e., at the $2-\sigma$ level). For these stars, whose exact 
number depends on the distance modulus adopted (see \S3.1.2 below), we do not attempt to 
assign individual mass values, since these would not be physically meaningful.


\subsection{Reliability Tests}

In order to ascertain the reliability of our adopted procedure, we have generated synthetic 
CMD's for the M3 HB, including synthetic photometric errors, with the purpose to verify 
whether our method is able to successfully recover a bimodal mass distribution, if one is 
present, given the uncertainties in the photometry and in the distance modulus. Since for 
the synthetic models the mass distribution is known a priori, this provides us with a crucial 
test of whether the mass distribution to be inferred from the actual observations can be 
trusted, insofar as the presence (or lack) of bimodality is concerned. 

In a blind experiment, one of us (M.C.) computed the synthetic distribution and shifted 
the derived distributions using a distance modulus/reddening combination not known to the 
other (A.V.)~-- who in turn attempted to infer the corresponding mass distribution (see below). 
Reassuringly, the distance modulus and reddening favored by A.V. in the process were in excellent 
agreement (i.e., to within 0.01~mag in both magnitude and color) with the input values, which 
in turn were very similar to those given in the Harris (1996) catalog, namely: $\mu_V=15.1$, 
$E(B\!-\!V) = 0.01$.

\subsubsection{Synthetic Distributions}
We have constructed four input bimodal distributions using the Monte Carlo code {\sc sintdelphi} 
(Catelan 2004 and references therein), as shown in the left panels of Figure~\ref{BimTest}. 
These mass distributions are closely patterned after the Castellani et al. (2005) proposed 
bimodal mass distribution for M3 (see \S1). Figure~\ref{BimTest} ({\em right panels})     
shows the recovered mass distributions, on the basis of our method, assuming the same 
distance modulus as used when constructing the synthetic distributions (i.e., $\mu_V=15.1$). 
While it is clear that the added synthetic errors     necessarily lead to some loss of 
information, and therefore to somewhat wider (inferred) mass distributions than in the 
synthetic models, our method always succeeds in recovering a bimodal distribution. Indeed, 
according to the KMM test (Ashman, Bird, \& Zepf 1994), the distributions shown on the 
right-hand panels of Figure~\ref{BimTest} are better described by bimodal rather than 
unimodal Gaussian distributions, with probability always higher than 99.99\%.

\subsubsection{Synthetic Distribution with Variable Distance Modulus}
    While the above results are suggestive, as we have seen there appears to be some 
uncertainty in the M3 distance modulus, which may also impact the derived mass distribution. 
We have investigated this source of systematic error by changing the adopted distance modulus 
over a wide range, from 15.0~mag to 15.4~mag, which should cover the full range of acceptable 
values (we recall that the value used in the simulations is 15.1~mag). In Figure~\ref{DMTest}, 
we show our results for three different input synthetic bimodal distributions ({\em upper panels}) 
for different values of the adopted distance modulus, from 15.0~mag ({\em second row}) to 15.4~mag 
({\em bottom row}). As can clearly be seen, while the choice of distance modulus does seem to 
affect the retrieved location of the peaks of the two output mass Gaussians, the bimodality in 
the mass distribution is always successfully recovered. 
Additional calculations and statistical tests using the KMM statistic 
confirm that we only (mistakenly) retrieve a unimodal mass distribution when the adopted 
distance modulus is in error by more than about 0.3~mag. Note that, when a distance modulus 
value that is less than the correct one by over 0.1~mag is assumed, we are unable to assign 
mass values to more than about $50\%$ of the stars, because many will then fall {\em below} 
the ZAHB (see the last paragraph preceding \S3.1). Naturally, such a loss of stars serves as an 
indicator that we have an incorrect distance modulus. In conclusion, our method seems to be 
very robust in its ability to detect mass bimodality among HB stars in globular clusters.

\subsubsection{Two Methods for the Determination of the Mass Distribution}\label{sec:3MET}
While the method we used to infer masses basically looks for the evolutionary track that 
goes closest to the actual data point in the CMD, other methods can be devised in which the 
evolutionary times are used as a diagnostic criterion. In particular, to avoid an excess of 
stars close to the helium exhaustion line, where evolution is very fast, one can 
alternatively choose the evolutionary track that goes through the 1-sigma error elipse 
with the smallest {\em evolutionary speed}, thus in practice giving higher weight to slower 
evolutionary stages. How would these different criteria affect the inferred mass 
distribution? 

In order to properly determine the evolutionary speed $v_{\rm evol}$, we first have to 
transform the CMD into a plane in which color and magnitude have comparable weights. This 
has been achieved using a technique similar to that described in Dixon et al. (1996), 
Catelan et al. (1998a), and Piotto et al. (1999), but here adapted to the $M_V$, $V\!-\!I$ 
plane. Accordingly, we define rescaled ``color'' $c$ and ``brightness'' $b$ coordinates 
as follows:  

\begin{eqnarray}
c & = & 168.3 \, (V\!-\!I)_0 + 96.64, \\
b & = & -42.67 \, M_V + 281.6. 
\end{eqnarray}

\noindent With this definition, $v_{\rm evol}$ is computed along an evolutionary track as

\begin{equation}
v_{\rm evol}= \frac{\sqrt{\Delta c^2 + \Delta b^2}}{\Delta t}. 
\end{equation}

Having thus defined the evolutionary speed, we have carried out numerical tests in which 
we generated, using {\sc sintdelphi}, a synthetic CMD with 1500 synthetic stars and an input 
mass distribution characterized by 
$\langle M\rangle = 0.642 \, M_{\odot}$, $\sigma_M = 0.020 \, M_{\odot}$. Synthetic error bars 
were added as in Catelan et al. (2001b). We then checked the impact of the different criteria 
upon the inferred mass distribution.  

In Figure~\ref{MDMethods} we compare the mass distributions that were inferred 
using the two different 
criteria indicated in the beginning of this section with the input distribution. In the 
{\em middle upper panel}, the evolutionary track passing closest to the actual data point was 
selected. In the {\em middle bottom panel}, the evolutionary track with the smallest 
evolutionary speed $v_{\rm evol}$ within the 1-sigma error ellipse was in 
turn selected. As can clearly be seen, the differences between the mass distributions 
derived using either of these two methods are very small, and they are both clearly 
able to properly reproduce the input mass distribution. On the other hand, the reader is 
warned that the method does not perform as nicely for more extreme HB types. This is also 
revealed by Figure~\ref{MDMethods}, where we show our attempts
to recover the input mass distribution in the cases of simulations computed for extremely 
blue {\em left panel} and red {\em right panel} HB morphologies. Irrespective of whether
we adopt the purely geometrical criterion or the criterion involving the minimum 
$v_{\rm evol}$ ({\em upper and bottom panels, respectively}), we are unable to 
reliably recover the input mass distribution in such cases. Photometry in other, more 
suitable bandpasses, where one finds a stronger dependence of colors and magnitudes on 
the stellar mass, would be required to achieve better results when the HB distribution 
is comprised of very blue or very red stars.  

We thus conclude that the simple method (in which the evolutionary track passing closest 
to the data point in the CMD is selected) appears to be good enough for our purposes, 
since it performs quite well for rather even, M3-like HB morphologies.

\section{Results}

   \begin{figure}[t]
   \centering
      \includegraphics[width=9cm]{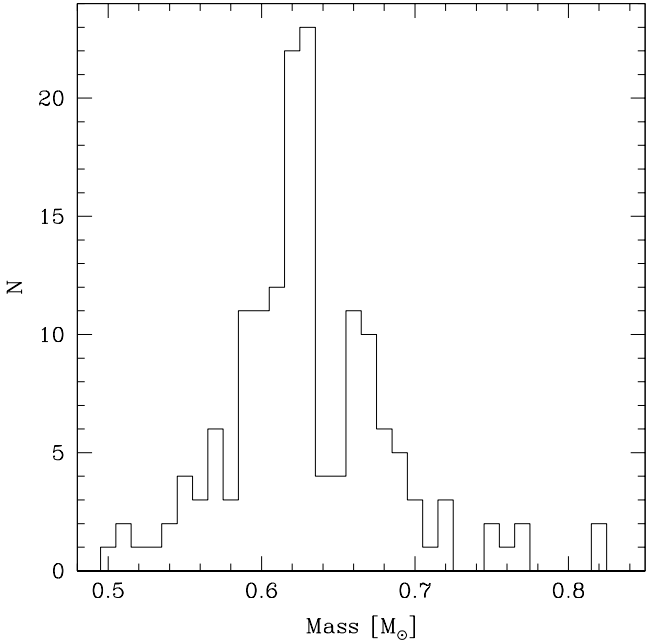}
      \caption{Mass distribution of nonvariable stars in the inner regions of M3. It displays 
	    two maxima, produced by stars pertaining to the red and blue HB domains. The minimum in 
		between indicates the existence of stars in highly evolved stages towards the end of the 
		HB phase.}
      \label{MDNoVar}
   \end{figure}

\subsection{Mass Distribution along the HB of M3}

The mass distribution of stars along the HB of M3 was obtained separately for the nonvariable 
and variable stars, since we had to rely on empirical data in the $VI$ bands for the former 
and in the $BV$ bands for the latter. Our final M3 mass distribution corresponds to the sum 
of these two, separately derived, distributions.

The non-variable stars that we selected for our study are those 
pertaining to the inner regions of M3. We inferred masses for 157 stars falling sufficiently 
close to HB evolutionary tracks, including 105 stars on the blue HB and 52 on the red HB 
({\em filled circles} in Fig.~\ref{CMDNoVar}). The stars plotted with open circles within 
the dashed region of Figure \ref{CMDNoVar} are those for which mass values were not assigned, 
since their photometry places them several $\sigma$ away from the predicted HB locus. The 
resulting mass distribution for these non-variable stars is bimodal, with peaks centered at 
$0.625\, M_\odot$ and $0.665\, M_\odot$ (Fig.~\ref{MDNoVar}). The two peaks reflect the masses 
of blue and red HB stars, respectively; naturally, we anticipate that the gap in between these 
two mass modes will be at least partially filled when due account is taken of M3's variable 
stars. Had we used the method described by Rood \& Crocker (1989), we would not have found 
any non-variable stars with masses around $0.640\, M_\odot$, since in that method the effects 
of evolution away from the ZAHB are not taken into account.

\begin{figure}[t]
   \centering
      \includegraphics[width=9cm]{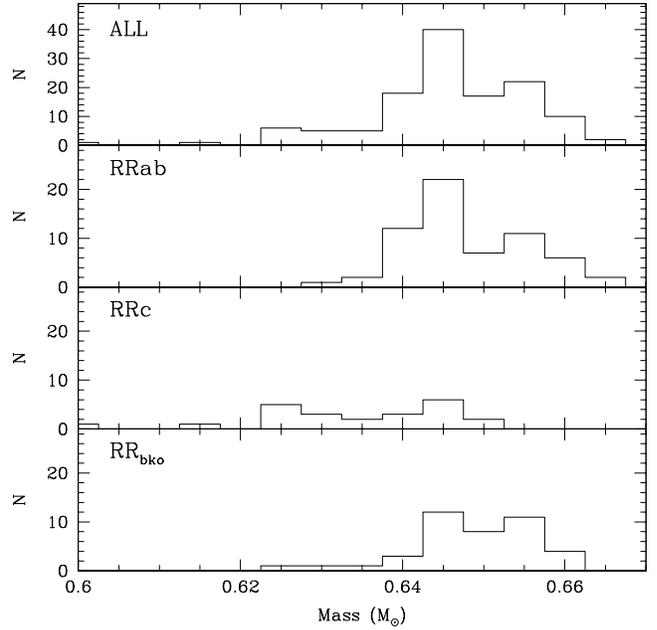}
      \caption{{\em Upper panel}: inferred mass distribution for all RR Lyrae variable stars 
	    in M3. The mass distributions for each RR Lyrae subtype are shown in the other panels: 
		RRab with well-behaved light curves ({\em second panel}), RRc ({\em third panel}), and 
		RR Lyrae showing the Blazhko effect ({\em bottom panel}).
      }
      \label{MDVar}
   \end{figure}

The mass distribution for variable stars (Fig.~\ref{MDVar}, {\em upper panel}) was obtained 
from 127 of the 133 stars that we chose initially. The remainder of the stars fall below 
the ZAHB, and their photometric errors do not allow them to be reconciled with the HB phase 
(some of them may be pre-ZAHB stars). A Gaussian fit provides a mean mass of $0.6445\, M_\odot$ 
and dispersion $0.0076\, M_\odot$. More in detail, a Gaussian fit to the mass distribution for 
the RRab stars has mean mass and dispersion given by $0.6440\, M_\odot$ and $0.0049\, M_\odot$, 
respectively (Fig.~\ref{MDVar}, {\em second panel}), whereas the Blazhko-type RR Lyrae have mean 
mass $0.6482\, M_\odot$ and dispersion $0.0085\, M_\odot$ (Fig.~\ref{MDVar}, {\em bottom panel}). 
For the RRc, in turn, a rather uniform mass distribution is inferred instead, though in this case 
the number of stars is much lower (Fig.~\ref{MDVar}, {\em third panel}). Note, in any case, that 
the derived masses for the RRc stars tend to be lower than for the RRab stars, which is consistent 
with the expectations of canonical theory (see, e.g., Fig.~\ref{CMDVar}). 

\begin{figure}[t]
   \centering
      \includegraphics[width=9cm]{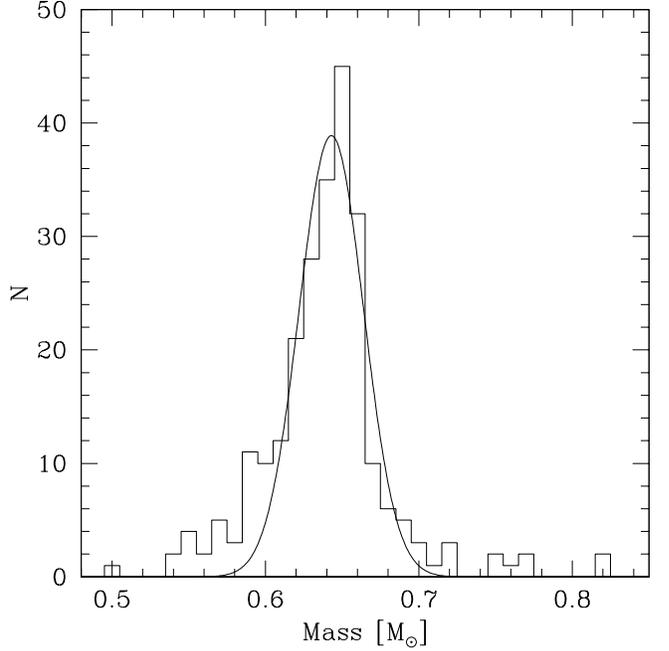}
      \caption{Mass distribution for the M3 HB, obtained by combining the mass distributions 
        for the variable and non-variable stars. The best-fitting Gaussian 
        ({\em solid line}) has $\left\langle M \right\rangle = 0.642\, M_\odot$ 
        and $\sigma = 0.020\, M_\odot$.}
      \label{MDTodas}
   \end{figure}

In order to study the {\em global} mass distribution along the M3 HB, we must suitably combine 
our samples of variable and non-variable stars. In particular, we must ensure consistency with 
the observed proportions of stars falling along the blue HB, the red HB, and inside the 
instability strip. We accordingly use the proportions 
$\mathcal{B:V:R}=39\!:\!40\!:\!21$ (Catelan 2004), where 
$\mathcal{B}$, $\mathcal{V}$, and $\mathcal{R}$ indicate the numbers of blue, variable, and red 
HB stars, respectively. As a consequence, given the sample of red HB stars in our study, 
in order to derive an unbiased, final mass distribution we must randomly remove 9 and 28 
stars from our blue HB and RR Lyrae samples, respectively. Our final derived mass distribution 
is shown in Figure~\ref{MDTodas}. 

We fit a Gaussian to this mass distribution, finding (Fig.~\ref{MDTodas}) 

\begin{equation}
\left\langle M \right\rangle = 0.642\, M_\odot, \,\,\,\,\, \sigma = 0.020\, M_\odot. 
\end{equation} 

\noindent This result is consistent with that previously derived by Rood \& Crocker (1989), 
who also find, using a method similar to ours but not taking into account evolutionary effects, 
a unimodal mass distribution, with $\left\langle M \right\rangle = 0.666\, M_\odot$ and 
$\sigma=0.018\, M_\odot$. Catelan et al. (2001b) have also found that a unimodal mass 
distribution along the M3 HB is consistent with the observed HB morphology parameters 
of the cluster: according to their results, the innermost cluster regions may be 
characterized by $\left\langle M \right\rangle = 0.637\, M_\odot$ and $\sigma=0.023\, M_\odot$, 
whereas the outermost regions, which seem to be a bit redder, may be described instead in terms 
of a normal distribution with $\left\langle M \right\rangle = 0.645\, M_\odot$ and 
$\sigma=0.018\, M_\odot$. 

On the other hand, while the fit shown in Figure~\ref{MDTodas} seems rather acceptable, 
close inspection reveals what appears to be an excess of stars on the wings of the distribution, 
especially at its low-mass end. That the true mass distribution may be unimodal but not precisely 
Gaussian is also supported by an analysis of its skewness and kurtosis (see \S14.1 in Press et al.
1992). We find values of ${\rm Skew} = 0.28\pm 0.16$ and ${\rm Kurt} = 3.31\pm 0.31$, whereas a 
perfect Gaussian has ${\rm Skew} \equiv 0$ and ${\rm Kurt} \equiv 3$.

\begin{figure}[t]
   \centering
      \includegraphics[width=9cm]{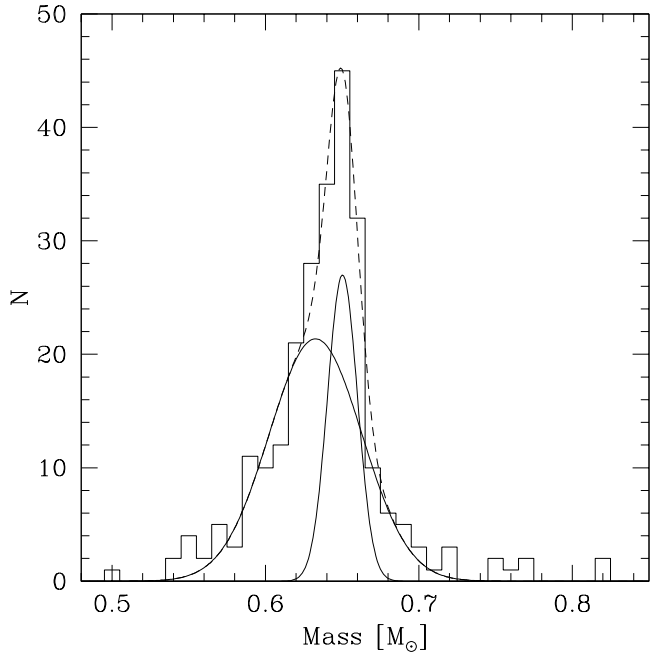}
      \caption{As in Figure~\ref{MDTodas}, but showing the best-fitting bimodal solution. 
        The two Gaussians ({\em solid lines}) have 
        $\left\langle M \right\rangle_1 = 0.633\, M_\odot$ and $\sigma_1 = 0.026\, M_\odot$
        (with 71.1\% of the stars). 
        and 
        $\left\langle M \right\rangle_2 = 0.650\, M_\odot$ and $\sigma_2 = 0.008\, M_\odot$
        (with 28.9\% of the stars). 
      }
      \label{MDTodas2}
   \end{figure}

Naturally, since the derived unimodal distribution is not perfectly Gaussian, an alternative 
description of the data can be accomplished with a linear combination of Gaussians. Indeed, 
when faced with the question to opt for a single Gaussian or a combination of two Gaussians, 
the KMM test gives, not surprisingly, preference to the latter, at a very high level of 
confidence ($>99.99\%$). The best-fitting bimodal solution is shown in Figure \ref{MDTodas2}, 
where the {\em dashed line} shows the result of the sum over the two derived Gaussians 
({\em solid lines}). In this case, the best-fitting Gaussians are given by 

\begin{equation}
\left\langle M \right\rangle_1 = 0.633\, M_\odot, \,\,\,\,\, \sigma_1 = 0.026\, M_\odot, 
\end{equation}

\begin{equation}
\left\langle M \right\rangle_2 = 0.650\, M_\odot, \,\,\,\,\, \sigma_2 = 0.008\, M_\odot,  
\end{equation}

\noindent with 71.1\% of the stars in the first mode and 28.9\% in the second. 

As previously discussed on the basis of Figures~\ref{BimTest} and \ref{DMTest}, one expects 
that the mass distribution determined in this way will differ somewhat from the ``intrinsic'' 
mass distribution of the cluster. We have made an effort to account for this effect, at least 
in the case of the bimodal solution, by comparing the derived and input mass modes, as given 
by the panels with $\mu_V = 15.1$ in Figure~\ref{DMTest}.  As a consequence, we infer that 
the ``true'' mass modes have on average $\sigma$ values which are only about 77\% of those 
given above. In like vein, the positions of the centers  of the high- and low-mass modes 
are on average shifted by $-0.008\,M_{\odot}$ and $+0.008\,M_{\odot}$, respectively. 
Therefore, the suggested (corrected) mass distribution for the M3 stars, assuming a 
bimodal solution, is the following:

\begin{equation}
\left\langle M \right\rangle_{1,\,{\rm cor}} = 0.625\, M_\odot, \,\,\,\,\, \sigma_{1,\,{\rm cor}} = 0.019\, M_\odot,
\end{equation}

\begin{equation}
\left\langle M \right\rangle_{2,\,{\rm cor}} = 0.658\, M_\odot, \,\,\,\,\, \sigma_{2,\,{\rm cor}} = 0.006\, M_\odot,  
\end{equation}

\noindent again with about 71\% of the stars in the low-mass mode and 29\% in the high-mass 
mode. 

Note that our derived mass distribution for M3 differs markedly from either the input or 
the derived solutions shown in Figures~\ref{BimTest} and \ref{DMTest}, since the two 
Gaussians here are much closer together than was derived in almost all distributions 
except in those with the largest distance moduli (which were off from the correct 
solution by 0.2~dex or more in $\mu_V$). Therefore, while our study may give some 
support to the existence of some degree of HB bimodality in M3, the derived distribution 
also differs in detail from the one that was set forth by Castellani et al. (2005). 
Interestingly, the two derived mass modes do not seem to be fully detached, contrary 
to what was suggested in the latter study. As a consequence, synthetic HB's based on 
the mass distribution that was inferred in this section are again unable to properly 
reproduce the observed fundamentalized period distribution in M3 (see \S1 for references
to prior work on this topic).   

We repeated the whole procedure described in this section, but 
using the set of tracks independently computed by Pietrinferni et al. (2004). As a 
result, we derived a mass distribution which is qualitatively very similar to that 
shown in Figures~\ref{MDTodas} and \ref{MDTodas2}. The only noteworthy difference is 
that, when using the Pietrinferni et al. tracks, the peak of the high-mass mode is 
slightly more pronounced than indicated in these plots.

 \begin{figure*}[t]
   \centering
      \includegraphics[width=15cm]{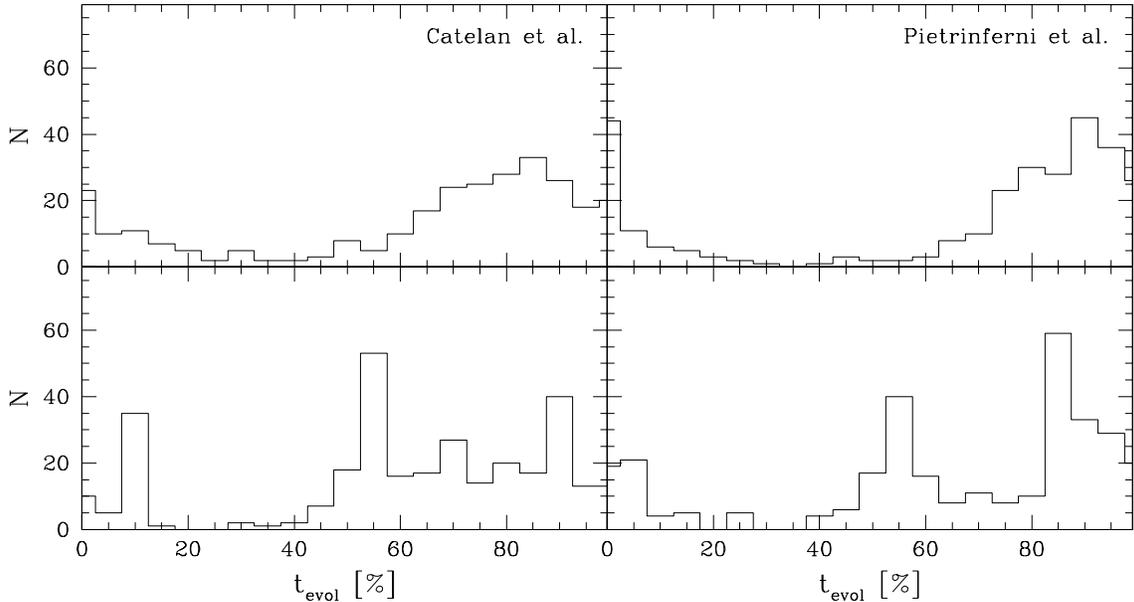}
      \caption{Histograms showing the derived distributions of evolutionary times 
	    $t_{\rm evol}$ for stars in the HB phase, in terms of fractions of 
	    the total HB lifetime for the star's inferred mass (where 0\% corresponds to 
		the ZAHB and 100\% to the TAHB). On the {\em left} we show the results obtained on
		the basis of our own evolutionary tracks, whereas on the {\em right} the results 
		based on the independent set of tracks by Pietrinferni et al. (2004) are 
		displayed. 	The {\em upper panels} show the evolutionary 
		times based on the HB tracks that go closest to each individual data point in 
		the CMD (e.g., those that would correspond to $t_{\rm evol} = 63.2\%$ and $100\%$ 
		for left and right panels in Figure~\ref{Elipses}, respectively). 
        The {\em bottom panels} correspond to the minimun evolutionary speed inside 
		the 1-sigma error ellipse (e.g., those that would correspond to 
		$t_{\rm evol} = 57.3\%$ and $99.8\%$ for the left and right panels in 
		Figure~\ref{Elipses}, respectively).		
      }
      \label{LifeTimes}
   \end{figure*}


\subsection{Evolutionary Times}
While we have previously argued (\S\ref{sec:3MET}) that the inferred mass 
distribution is fairly robust in regard to the different recipes for using 
the evolutionary lifetimes to infer the mass values, we have so far not 
discussed what happens to the resulting distribution of evolutionary 
lifetimes itself. In principle, assuming a smooth feeding of the HB phase
from the RGB tip, one should have an essentially flat distribution between 
0 and 100\% of the HB lifetime, except of course for statistical fluctuations. 
How does the inferred distribution of evolutionary times look like, in the 
case of M3? 

The answer is provided in Figure~\ref{LifeTimes} ({\em left}). 
In this plot, the {\em upper panel} shows the derived $t_{\rm evol}$ 
distribution, corresponding to the 
mass distribution that was inferred by simply adopting the evolutionary 
tracks that go closest to each individual data point in the CMD. As can 
clearly be seen, there appears to be a dearth of stars in the range 
$t_{\rm evol} \approx 20-60\%$, and/or an excess of stars with 
$t_{\rm evol} \approx 70-90\%$. The {\em lower panel} in the same figure 
reveals that, by adopting the smallest $v_{\rm evol}$ value inside the 
1-sigma error ellipse for each data point, not only is the problem not solved, 
but also a large excess of stars with $t_{\rm evol} \approx 10\%$ results 
as well, as well as another strong peak of stars with 
$t_{\rm evol} \approx 55\%$. Figure~\ref{LifeTimes} ({\em right}) shows 
the result of the same exercise, performed using the Pietrinferni et al. 
(2004) evolutionary tracks. As can clearly be seen, the results are 
qualitatively very similar, thus indicating that the problem does not 
lie in our specific choice of evolutionary tracks. 

As a matter of fact, the presence of these two peaks is an expected, direct 
consequence of the minimum $v_{\rm evol}$ method. As well known, HB evolution 
is slowest close to the ``turning points'' on the CMD, including the blue 
nose and red ``noses'' that are clearly seen in the evolutionary track 
displayed in Figure~\ref{CMDVar}. Therefore, any method based on minimum 
$v_{\rm evol}$ values will tend to preferentially pick the $t_{\rm evol}$
values associated to these features~-- as in the case of Figure~\ref{Elipses} 
({\em left panel}), where clearly the ``blue nose'' position was selected 
for a star whose original position on the CMD implied a slightly more 
advanced evolutionary stage. We conclude that the peak at 
$t_{\rm evol} \approx 55\%$ can be ascribed to these ``blue noses,'' 
whereas the peak at $t_{\rm evol} \approx 10\%$ is instead due to the 
``red noses'' that occur slightly after the star reaches the ZAHB. 

\begin{figure}[t]
   \centering
      \includegraphics[width=9cm]{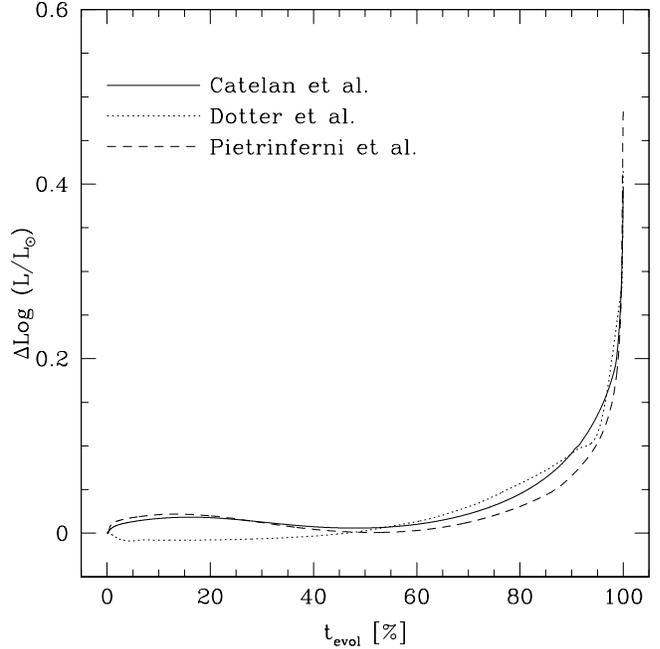}
      \caption{Luminosity evolution from the ZAHB, 
	  $\Delta\log(L/L_{\odot}) \equiv \log L(t_{\rm evol}) - \log L (t_{\rm evol}=0)$, 
	  for three independent sets of evolutionary tracks. Evolutionary tracks whose 
	  ZAHB position lies close to the middle of the instability strip is shown 
	  in all cases. As can clearly be seen, according to the models one should  
	  expect $\approx 65-70\%$ of all stars along the horizontal part of the HB to lie 
	  within 0.05~mag of the ZAHB. Observations indicate instead that only 
	  $\approx 40\%$ are found within this range (see text). 
      }
      \label{fig:lumevol}
   \end{figure}

Irrespective of the method adopted, the main problem revealed by 
Figure~\ref{LifeTimes} is the fact that there are many fewer HB stars 
with $0\% \leq t_{\rm evol} \leq 50\%$ than there are stars with 
$50\% < t_{\rm evol} \leq 100\%$. 
We have checked that this is not a problem affecting only a group of stars 
along M3's HB, but in fact the $t_{\rm evol}$ distribution looks bimodal 
for red HB, blue HB, and RR Lyrae stars. At present, we do not have an 
explanation for this problem, other than speculating that the present 
set of evolutionary tracks predicts too little luminosity evolution, 
thus leading to too few predicted stars at high luminosities, compared 
to the observations (see also \S2.2.4 in Catelan et al. 2001a, and 
\S3.2 in Catelan et al. 2001b). We have checked that this is not a problem 
referring exclusively to our adopted models; in fact, similar (or even 
more extreme) discrepancies are suggested by the Pietrinferni et al. 
(2004) or the Dotter et al. (2007) evolutionary tracks. This is further 
illustrated in Figure~\ref{fig:lumevol}, which shows that, irrespective 
of the set of models used, HB stars are predicted to spend too little time 
at relatively 
high luminosities, compared with what is suggested by the observed CMD. 
More specifically, along the horizontal part of the HB, one expects to 
find, according to these models, $\approx 65-70\%$ of the stars within about 
0.05~mag of the ZAHB. The observations, on
the other hand, reveal $\approx 40\%$ of the HB stars within such a 
magnitude range from the ZAHB, even including in the tally those stars 
that fall below the ZAHB (presumably due to photometric errors; see, e.g.,  
Fig.~\ref{CMDVar}). If we changed the adopted ZAHB position so as to better 
accomodate these ``sub-ZAHB stars,'' the noted discrepancy would become even 
more dramatic. 

Naturally, until these problems are conclusively solved, mass distributions 
based on the CMD method, such as the one provided in the present paper, 
should be considered tentative. In like vein, we caution that the fact that 
there are also too many (presumably) evolved {\em red} HB stars argues against 
a solution to this specific problem based on a component with a high helium 
abundance among the blue HB stars.

\section{Summary and Conclusions}
In the present paper, we have studied the mass distribution of M3's HB stars by 
comparing their observed locations in the CMD with the predictions of evolutionary 
models. Our results suggest that, within the canonical framework, M3's 
HB mass distribution can be characterized either by a unimodal mass distribution that 
is not perfectly Gaussian or by a bimodal mass distribution in which the two mass modes 
are adequately described by Gaussians. In the latter case, the two mass modes appear to 
be separated in mean mass by only $\approx 0.03\,M_{\odot}$, whereas the dispersion in 
mass of the low-mass mode is significantly higher than that of the high-mass mode. 

As far as the distribution of evolutionary times is concerned, we find that it is 
not in good agreement with the canonical expectations, in that we obtain a bimodal 
distribution, with a dearth of stars with $t_{\rm evol} \approx 20-60\%$ and/or an 
excess of stars with $t_{\rm evol} \approx 70-90\%$. This suggests that the present 
evolutionary models underestimate the luminosity evolution along the HB. We have 
checked that other sets of evolutionary tracks, such as the ones computed by 
Pietrinferni et al. (2004) or Dotter et al. (2007), indicate similar, or even 
smaller, luminosity evolution than the present ones. Until this problem is solved 
and better agreement between observed and predicted $t_{\rm evol}$ values can be achieved, 
mass distributions derived using a CMD-based method, such as the one presented in 
this paper, should be considered tentative. Conversely, no  solution for the HB star 
problems in M3 that were previously discussed by several authors (Rood \& Crocker 
1989; Catelan 2004; Castellani et al. 2005; D'Antona \& Caloi 2008) can be 
considered complete until good agreement between predicted and observed 
lifetime distributions is finally achieved (see also Catelan et al. 2001a,b).    


\begin{acknowledgements}
Support for M.C. is provided by Proyecto Fondecyt Regular No. 1071002. We thank 
M. Zoccali for interesting discussions, 
and an anonymous referee for her/his comments which have led to a significantly 
improved manuscript. 

Cristi\'an Aruta, in memoriam. 
\end{acknowledgements}

\begin{appendix}

\section{Photometry in the $BV$ Bands: Is there a Problem?}

\begin{figure}
   \centering
      \includegraphics[width=9cm]{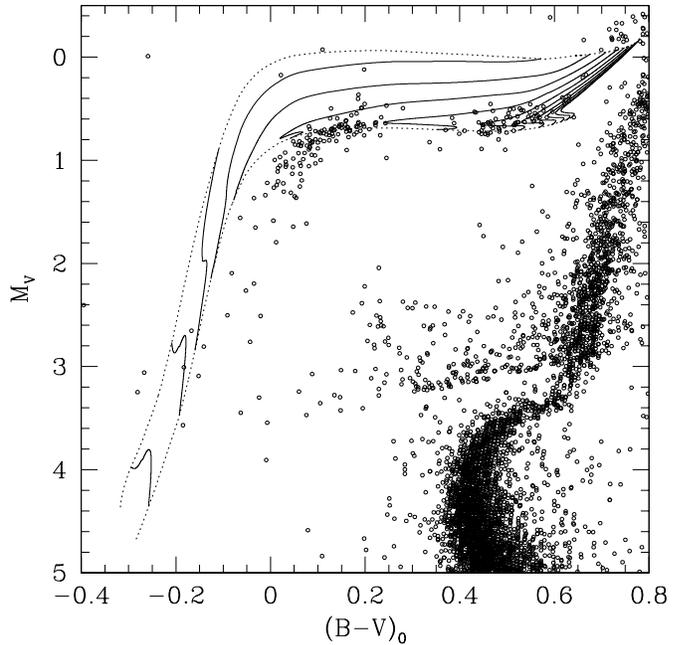}
      \caption{$M_V$, $B\!-\!V$ CMD of nonvariable stars in the outer regions of M3. Evolutionary tracks ({\em solid lines}), along with the corresponding ZAHB and TAHB lines ({\em dotted lines}), are overplotted. As can clearly be seen, for colors
      bluer than $(B\!-\!V)_0 = 0.2$ the models progressively deviate from the empirical data. As discussed in the main text, this is likely due to an error in the calibration of the photographic data [compare with Fig.~1, which reveals excellent agreement between the models and the CCD data in the $M_V$, $(V\!-\!I)_0$ plane].
      }
      \label{FotoErr}
   \end{figure}

In Figure~\ref{FotoErr} we overplot our evolutionary tracks with the observational data for the outer regions of M3. As can be seen, there is a clear disagreement between the predicted locus of M3's blue HB stars and the observations. We find no combination of $\mu_V$ and $E(B\!-\!V)$ that allows us to provide good agreement between the models and the empirical data. On the other hand, for the inner regions of M3 (Fig.~\ref{CMDNoVar}) no such problem was present. Since the measurements for the inner regions are based on CCD data (Ferraro et al. 1997) whereas for the outer regions photographic data were used (Buonanno et al. 1994), this strongly suggests an error in the calibration of the photographic data, especially in regard to the color term. This is consistent with the discussion in Ferraro et al., who have also called attention to the possibility of large errors in the $B$-band photometry. Note that this problem does {\em not} affect the adopted colors and magnitudes for the RR Lyrae stars, 
whose adopted $BV$ photometry (\S2) came from an independent source (namely, Cacciari et al. 2005 and references therein). 

\end{appendix}


\begin{thebibliography}{}
 
\bibitem[Ashman et al.(1994)Ashman, Bird, \& Zepf]{kaea94}
  Ashman, K. M., Bird, C. M., \& Zepf, S. E. 1994, \aj, 108, 2348
 
\bibitem[Bailey(1913)]{sb13}
  Bailey, S. I. 1913, Harv. Coll. Observ. Annals, 78, 1

\bibitem[1995]{Bono} 
  Bono, G., Caputo, F., \& Stellingwerf, R. F. 1995, \apjs, 99, 263
 
\bibitem[1994]{Buonanno} 
  Buonanno, R., Corsi, C. E., Buzzoni, A., Cacciari, C., Ferraro, F. R., 
    \& Fusi Pecci, F. 1994, \aap, 290, 69

\bibitem[2005]{Cacciari} 
  Cacciari, C., Corwin, T. M., \& Carney, B. W. 2005, \aj, 129, 267

\bibitem[Caloi et al.(1978)Caloi, Castellani, \& Tornamb\`e]{vcea78}
  Caloi, V., Castellani, V., \& Tornamb\`e, A. 1978, \aaps, 33, 169 

\bibitem[Caloi \& D'Antona(2007)]{cd07}
  Caloi, V., \& D'Antona, F. 2007, \aap, 463, 949 

\bibitem[Caloi \& D'Antona(2008)]{cd08}
  Caloi, V., \& D'Antona, F. 2008, \apj, in press (astro-ph/0709.1572)  

 \bibitem[2005]{Castellani} 
  Castellani, M., Castellani, V., \& Cassisi, S., 2005, \aap, 437, 1017

 \bibitem[2004]{Catelan04} 
  Catelan, M. 2004, \apj, 600, 409

\bibitem[Catelan et al.(2001a)]{cea01a}
  Catelan, M., Bellazzini, M., Landsman, W. B., Ferraro, F. R., Fusi Pecci, F., 
  \& Galleti, S. 2001a, \aj, 122, 3171

 \bibitem[1998a]{Catelan98a} 
  Catelan, M., Borissova, J., Sweigart, A. V., \& Spassova, N. 1998a, \apj, 494, 265

 \bibitem[2001b]{Catelan01b} 
  Catelan, M., Ferraro, F. R., \& Rood, R.T. 2001b, \apj, 560, 970

\bibitem[Catelan et al.(2004)Catelan, Pritzl, \& Smith]{mcea04} 
  Catelan, M., Pritzl, B.~J., \& Smith, H.~A.\ 2004, \apjs, 154, 633 

\bibitem[Catelan et al.(1998b)Catelan, Sweigart, \& Borissova]{mcea98b}
  Catelan, M., Sweigart, A. V., \& Borissova, J. 1998b, in A Half Century of Stellar 
    Pulsation Interpretation, ASP Conf. Ser., Vol. 135, ed. P. A. Bradley \& J. A. 
	Guzik (San Francisco: ASP), 41
  
\bibitem[Clementini et al.(2004)]{gcea04}
  Clementini, G., Corwin, T. M., Carney, B. W., \& Sumerel, A. N. 2004, \aj, 127, 938

\bibitem[Corwin \& Carney(2001)]{cc01}
  Corwin, T. M., \& Carney, B. W. 2001, \aj, 122, 3183


\bibitem[Crocker et al.(1988)Crocker, Rood, \& O'Connell]{dcea88}
  Crocker, D. A., Rood, R. T., \& O'Connell, R. W. 1988, \apj, 332, 236

\bibitem[Dixon et al.(1996)]{vdea96}
  Dixon, W. V. D., Davidsen, A. F., Dorman, B., \& Ferguson, H. C. 1996, \aj, 
  111, 1936

\bibitem[Dotter et al.(2007)]{adea07}
  Dotter, A., Chaboyer, B., Jevremovi\'{c}, D., Baron, E., Ferguson, J. W., Sarajedini, A., 
    \& Anderson, J. 2007, \aj, 134, 376  
  
 \bibitem[1997]{Ferraro} Ferraro, F. R., Carretta, E., Corsi, C. E., FusiPecci, F., Cacciari, C., 
 Buonanno, R., Paltrinieri, B., \& Hamilton, D. 1997, \aap, 320, 757

 
 \bibitem[1996]{Harris} Harris, W. E. 1996, \aj, 112, 1487

 \bibitem[1982]{Hill} Hill, G. 1982, Publ. Dom. Astrophys. Obs., 16, 67
 
\bibitem[Kaluzny et al.(1998)]{jkea98}
  Kaluzny, J., Hilditch, R. W., Clement, C., \& Rucinski, S. M. 1998, \mnras, 296, 347
 
\bibitem[Lee et al.(1990)Lee, Demarque, \& Zinn]{ldz90}
  Lee, Y.-W., Demarque, P., \& Zinn, R. 1990, \apj, 350, 155
 
 \bibitem[2003]{Marconi} Marconi, M., Caputo, F., Di Criscienzo, M., \& Castellani, M. 2003, 
 \apj, 596, 299

 
\bibitem[Oosterhoff(1939)]{o39}
  Oosterhoff, P. Th. 1939, Observatory, 62, 104

\bibitem[Oosterhoff(1944)]{o44}
  Oosterhoff, P. Th. 1944, Bull. Astron. Inst. Neth., 10, 55

\bibitem[Pietrinferni et al.(2004)]{apea04}
  Pietrinferni, A., Cassisi, S., Salaris, M., \& Castelli, F. 2004, \apj, 612, 168

\bibitem[Piotto et al.(2007)]{gpea07}
  Piotto, G., et al. 2007, \apjl, 661, L53

\bibitem[Piotto et al.(1999)]{gpea99}
  Piotto, G., Zoccali, M., King, I. R., Djorgovski, S. G., Sosin, C., Rich, R. M., 
  \& Meylan, G. 1999, \aj, 118, 1727

\bibitem[Press et al.(1992)]{wpea92}
  Press, W. H., Teukolsky, S. A., Vetterling, W. T., \& Flannery, B. P. 1992, Numerical Recipes 
    in C (Cambridge: Cambridge Univ. Press)

\bibitem[Rieke \& Lebofsky(1985)]{gr85} 
  Rieke, G. H., \& Lebofsky, M. J. 1985, \apj, 288, 618

\bibitem[Roberts \& Sandage(1995)]{rs95}
  Roberts, M. S., \& Sandage, A. 1995, \aj, 60, 185

\bibitem[Rood(1973)]{rtr73}
  Rood, R. T. 1973, \apj, 184, 815

 \bibitem[1989]{RC89} Rood R. T., \& Croker, D. A. 1989, in IAU Colloq. 111, 
 The Use of Pulsating Stars in Fundamental Problems of Astronomy, ed. E. G. Schmidt 
 (Cambridge: Cambridge Univ. Press), 103 

\bibitem[Schlegel, Finkbeiner, \& Davis(1998)Schlegel et al.]{dsea98}
  Schlegel, D. J., Finkbeiner, D. P., \& Davis, M. 1998, \apj, 500, 525

\bibitem[Smith(1995)]{hs95}
  Smith, H. A. 1995, RR Lyrae Stars (Cambridge: Cambridge Univ. Press) 

\bibitem[Sneden et al.(2004)]{csea04}
  Sneden, C., Kraft, R. P., Guhathakurta, P., Peterson, R. C., \& Fulbright, J. P. 
    2004, \aj, 127, 2162  

\bibitem[Szeidl(1973)]{bs73}
  Szeidl, B. 1973, Comm. Konk. Obs., 63, 1

 
 \bibitem[2003]{VanderBerg} VandenBerg, D. A., \& Clem, J. L. 2003, \aj, 126, 778      

\end{thebibliography}
\end{document}